\documentclass[superscriptaddress,showpacs,amsmath,amssymb,pra,preprint]{revtex4}
\usepackage[dvips]{graphicx}
\usepackage{psfrag}

\begin{document}

\title{\bf The barriers to producing multiparticle superposition states in rotating Bose-Einstein condensates}
\label{sec-cat}
\author{David W. Hallwood and Keith Burnett}
\affiliation{Clarendon Laboratory, University of Oxford, Parks Road, Oxford OX1
3PU, United Kingdom} 
\author{Jacob Dunningham}
\affiliation{School of Physics and Astronomy, University of Leeds, Leeds LS2 9JT,
United Kingdom} 

\begin{abstract}
We study why it is quite so hard to make a superposition of superfluid flows in a Bose-Einstein condensate. To do this we initially investigate the quantum states of $N$ atoms trapped in a 1D ring with a barrier at one position and a phase applied around it. We show how macroscopic superpositions can in principle be produced and investigate factors which affect the superposition. We then use the Bose-Hubbard model to study an array of Bose-Einstein condensates trapped in optical potentials and coupled to one another to form a ring. We derive analytic expressions for the quality of the superposition for this system, which agrees well with direct diagonalisation of the Hamiltonian for relatively small numbers of atoms. We show that for macroscopic superpositions to be realised there are essentially three straightforward requirements, other than an absence of decoherence, which become harder to achieve as the system size increases. Firstly, the energies of the two distinct superfluid states must be sufficiently close. Secondly, coupling between the two states must be sufficiently strong, and thirdly, other states must be well separated from those participating in the superposition.
\end{abstract}


\maketitle

\section{Introduction}

One of the most puzzling questions in physics concerns why we do not see quantum superposition phenomena in the macro world~\cite{leggett_02}. In quantum mechanics superposition states are a crucial part of the description of the micro world, while in classical mechanics they are meaningless. This was highlighted by the thought experiment described by Schr\"{o}dinger in 1935, where a cat is put into a superposition of alive and dead states~\cite{schrodinger_35}. Although Schr\"{o}dinger's thought experiment is unachievable, there have been many more realistic proposals for creating systems that can be placed in a macroscopic superposition if the system is sufficiently decoupled from the environment~\cite{caldeira_81,leggett_87}. In this paper we focus on what prevents the production of a cat state in superfluid flow, rather than on decoherence issues. Our discussion leads to some rather straightforward but nevertheless important conclusions.

Superpositions have been experimentally studied in various forms~\cite{tonomura_02}, including two slit electron interference~\cite{merli_76,tonomura_89} and interferometry with atoms and molecules~\cite{carnal_91,Keith_91,Noel_95,arndt_99,brezger_02}. More recently larger systems have also been put into superposition states~\cite{lee_05,leibfried_05,Mitchell2004a,ourjoumtsev_06,Arndt1999a,Sackett2000a}. Experimental signatures of large scale quantum superpositions have been observed in superconducting quantum interference devices (SQUIDS). The superposition was between states of different trapped flux or opposite currents flowing around the device. Truly macroscopic systems can be defined as being visible with the naked eye or have a macroscopic measurable quantity associated with them. The currents measured in the SQUID consisted of approximately $10^9$ Cooper pairs in a loop of diameter $140 \, \mu\textrm{m}$ and produce a measurable magnetic flux, meaning tunnelling between two macroscopically distinct states had been achieved~\cite{rouse_95,friedman_00, wal_00}.  

BECs hold promise for realising similar results as they contain $10^3 - 10^7$ atoms with a high proportion in the same quantum state. They also have significant advantages over SQUIDs since they are highly controllable: the coupling between condensates and the strength of the interactions between atoms can be tuned over many orders of magnitude. They are also weakly interacting, which allows us to develop simple models to investigate macroscopic quantum effects in BECs at the microscopic level~\cite{jaksch_98}. To study macroscopic superpositions we need good control of the macroscopic parameters that influence the energy of the quantum states of the system. In the case of a SQUID the macroscopic variable is the external magnetic field, which can be applied accurately to the system over a large range of values. This allows an easily controllable phase to be applied around the loop. By Larmor's theorem~\cite{rosenfeld_65}, the analogue of applying an external magnetic field to charged particles in a loop is to move to the rest frame of the rotating potential that holds the neutral atoms of the BEC. The same effect could also be achieved by producing a flow of atoms round the loop by using Bragg scattering to imprint phases on the lattice sites~\cite{saba_05,shin_05}, or producing an effective magnetic field using two resonant laser beams~\cite{jaksch_03,juzeliunas_05}.

There have already been a number of theoretical proposals for creating cat states and several using Bose-Einstein condensates (BECs)~\cite{leggett_02,leggett_87a,cirac_98,gordon_99,ruostekoski_98,dunningham_01,dunningham_06,hallwood_06}. In section~\ref{sec:loop} we first look at a simple 1-D loop with $N$ particles on it and a barrier placed at one position. Using perturbation theory we show how a cat state of quasi-momentum (or flow) can be produced, and investigate the effects that inhibit its realisation. This analysis is developed in section~\ref{sec:BEC} using the Bose-Hubbard model (BHM) approach developed by Hallwood \emph{et al.}~\cite{hallwood_06}. It enables us to study quantum effects on an atomic level using the BHM~\cite{jaksch_98} rather than the macroscopic approach used to describe the SQUID~\cite{friedman_00, wal_00,leggett_02,leggett_87a} . The problem is tackled using the perturbation results derived in section~\ref{sec:loop} for an effective two state system. The original $S \times S$ matrix Hamiltonian, where $S$ is the total number of states, is reduced to the desired $2 \times 2$ matrix leaving a single coupling term between the two distinct multiparticle states where the atoms share the same flow state. This is done by finding all the possible coupling paths between the two multiparticle states via intermediate states. From this model we are able to understand what factors inhibit the creation of a superposition as the system gets larger even if decoherence effects can be removed. In order to try and improve the ``quality'' of the superposition we consider differences between our model and that of the SQUID in section~\ref{sec:LRF}. Our analysis supports that presented by Leggett in ref.~\cite{leggett_87a} where he emphasises the role of the adiabatic limit in the operation of a SQUID.

\section{Loop model}
\label{sec:loop}
When discussing macroscopic systems we need to be careful with our terminology. In this paper we refer to a \emph{flow state} as a state that describes the quasi-momentum or flow of each atom, $\Psi_{k_1,k_2,...,k_N}(r_1,r_2,...,r_N)$, where $r_i$ and $k_i$ represent the position and quantised value of flow for atom $i$. The states obtained by diagonalising a Hamiltonian are called \emph{energy states} with the \emph{ground state} having the lowest energy. States with all atoms with the same flow are called \emph{single flow states} and the other states where not all atoms have the same flow are called \emph{multiple flow states}.

A cat state takes the form,
\begin{equation}
\Psi(r_1,r_2,...,r_N) = \frac{1}{\sqrt{2}} \big(\Psi_0(r_1,r_2,...,r_N) + e^{i \theta}\Psi_1(r_1,r_2,...,r_N)\big)/\sqrt{2} \label{eq:sup},
\end{equation}
where $\Psi_0(r_1,r_2,...,r_N)$ and $\Psi_1(r_1,r_2,...,r_N)$ are both single flow states containing $N$ particles. In these states all particles have the same flow, so only one flow is labelled. The most general description for a system of particles is the many-body Schr\"{o}dinger equation, which has the form,
\begin{equation}
i\hbar\frac{\partial}{\partial t}\Psi_{k_1k_2...}(r_1,r_2,...) = \Bigg[ -\frac{1}{2} \sum_{i=1}^N \frac{\hbar^2}{m_i} \nabla_{r_i}^2 + \sum_{i<j} V_I( r_i-r_j) + \sum_i V(r_i) \Bigg] \Psi_{k_1k_2...}(r_1,r_2,...). \label{eq:MBSE}
\end{equation}
Here $V_I( r_i-r_j)$ is the interatomic potential and $V(r_i)$ is the external potential. 

\begin{figure}
\psfrag{phi}{$\phi$} \psfrag{R}{$R$} \psfrag{x}{$x$}
\psfrag{Barrier}{Barrier}
\includegraphics[width=4cm]{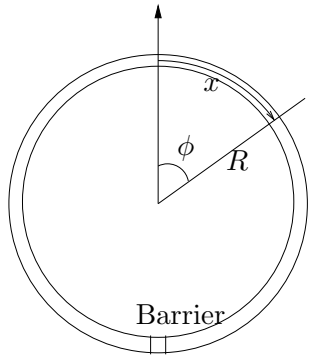}
\caption{\small Shows the system to be simulated. The particles are constrained to a loop of radius $R$ and has a barrier centred at $\phi = \pi$.}
\label{fig:loop}
\end{figure}

We cannot solve Eq.~\ref{eq:MBSE} directly, so we first simplify and then use perturbation theory to investigate some relevant features. Initially, we consider one particle confined in a one dimensional loop of radius $R$. There is only an angular dependence, $\theta$, or if we wish to consider the position on the circumference, an $x = R \theta$ dependence. Eventually we will apply a phase around the loop, which can be considered as a rotation of the loop, then use perturbation theory to consider the effect of a barrier (see Fig.~\ref{fig:loop}). We then consider state with several particles. 

If all interactions and barriers are ignored the Schr\"{o}dinger equation takes the simple free particle form:
\begin{equation}
-\frac{\hbar^2}{2 m}\frac{\partial^2}{\partial x^2}\psi_{k_n}(x) = E_{k_n} \psi_{k_n}(x).
\end{equation}
which has solutions $\psi_{k_n}(x) = \aleph e^{i 2 \pi {k_n}  x / L}$, and energy $E_{k_n}= \frac{\hbar^2}{2 m} \left( \frac{2 \pi}{L} \right) ^2 {k_n}^2$, where $\aleph = 1 / \sqrt{L}$ is a normalisation constant, $L = 2 \pi R$ is the circumference of the loop and $k_n$ is an integer representing the different flow states. For example, when $k_n = 0$ there is zero flow, and when $k_n = 1$ there is a $2 \pi$ phase around the loop indicating one quantum of clockwise flow. Throughout this paper we shall use the convention that positive phase variations correspond to clockwise flow. The quantisation of flow is due to the periodic boundary conditions of a loop.

Applying a phase around the loop changes the momentum term in the Hamiltonian. This is similar to the vector potential term added to the Hamiltonian of an atom in an external magnetic field or moving to the rest frame of a rotating object~\cite{rosenfeld_65}. For instance if we rotated the loop and moved into its rest frame the momentum of an atom becomes ${\bf p}_x \rightarrow {\bf p}_x - m {\bf v}_x$, where ${\bf p}_x$ is the momentum of the atom in the lab frame and ${\bf v}_x = R d \theta / d t$ is the velocity that the loop is rotating in the lab frame. The additional velocity can be represented in terms of phase,
\begin{equation}
{\bf v}_x = (\hbar / m) (\phi / L) \label{eq:appl_ph}
\end{equation}
where $\phi$ is the applied phase to the loop and $L$ is the circumference of the loop. We can see where Eq.~\ref{eq:appl_ph} comes from by considering the macroscopic wave function of a BEC~\cite{rey_03}, $\psi (x) =\psi_0e^{i \Phi (x)}$, where $\Phi (x)$ is the phase of the condensate and $\psi_0$ is the density, which we assume is constant for all $x$. When the phase is not constant throughout the condensate there is a velocity field associated with it,
\begin{eqnarray}
{\bf v} &=& \langle\psi(x)|{\bf p}/m|\psi(x)\rangle
            =-i\frac{\hbar}{m}\int\psi^*(x)\nabla\psi(x)dx \nonumber \\
           &=& \frac{\hbar}{m}\langle\nabla \Phi(x)\rangle.
\end{eqnarray}
If we move to the rest frame of a rotating loop then there will be an effective phase around the loop, $\phi$, in addition to the phase due to the momentum of the atoms, $\Phi$. By considering a linearly varying phase around the loop we get Eq.~\ref{eq:appl_ph}. In the rotating frame the Schr\"{o}dinger equation now takes the form:
\begin{equation}
\frac{({\bf p}_{x} - \hbar \phi / L)^2}{2 m}\psi_{k_n}(x) = E_{k_n} \psi_{k_n}(x).\label{eq:rot_schrd}
\end{equation}

\begin{figure}
\psfrag{energy}{Energy$/C$} \psfrag{phase}{Phase $\phi$} \psfrag{pi}{$\pi$} \psfrag{pi2}{$\pi/2$} \psfrag{3pi2}{$3\pi/2$} \psfrag{2pi}{$2\pi$}
\includegraphics[width=8cm]{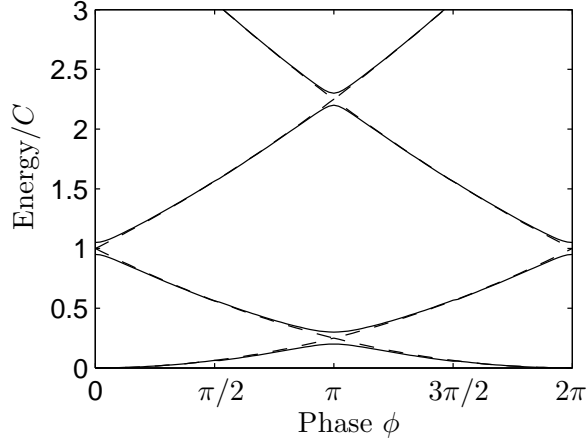}
\caption{\small Shows how the energy levels are affected by the phase around the loop. The dashed lines show a loop with no barriers, while the solid line shows a loop with a small barrier. $C = \frac{\hbar^2}{2 m} \left( \frac{2 \pi}{L}  \right) ^2$ is a constant and a constant energy term has been subtracted from the solid lines so the graphs line up.}
\label{fig:spectrum}
\end{figure}

Even though the system is finite in size, an atom travelling around the loop will see a periodic potential, so the system is in effect a perfect infinite periodic potential. The solution of this equation is the Bloch function~\cite{ashcroft_76}, $\Psi_K(x) = \exp[iKx]u_K(x)$, where $K$ is the Bloch wave vector (or wave number in one-dimension). This produces the energy in the form,
\begin{equation}
E_{k_n}(\phi) = \frac{\hbar^2}{2 m} \left( \frac{2 \pi}{L} \right) ^2 ({k_n} - \frac{\phi}{2 \pi})^2.
\end{equation}
where in our model $K=-\phi / L$. The energy levels are shown in Fig.~\ref{fig:spectrum}. The dashed lines show the calculated energy levels, where the zero flow state has zero energy at $\phi=0$, the one flow state has zero energy at $\phi=2\pi$, etc... The solid lines show the energy levels when the ring has a barrier at some point. A constant energy term has been removed so the graphs line up.

\begin{figure}
\psfrag{eps}{$\varepsilon$}\psfrag{E0}{$E^0$}\psfrag{E1}{$E^0_0$}\psfrag{E2}{$E^0_1$}\psfrag{E1'}{$E^1_0$}\psfrag{E2'}{$E^1_1$}\psfrag{phi}{$\phi$}\psfrag{Energy}{Energy}
\includegraphics[width=8cm]{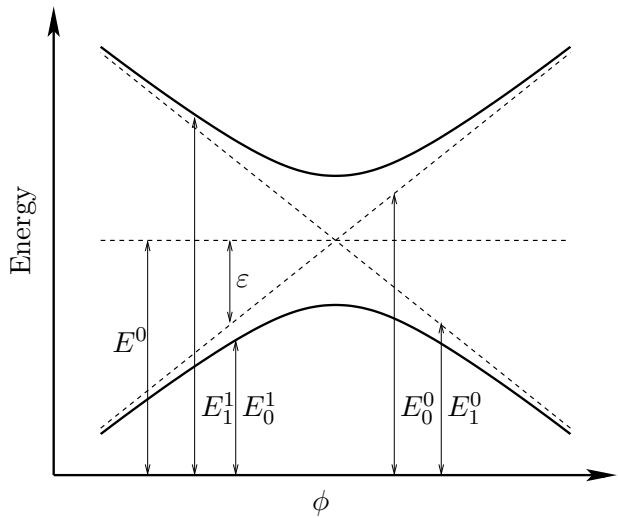}
\caption{\small Shows a 2 level system where the energy levels have an anti-crossing at $\phi = \pi$. $E^0$ is the energy at the crossing, $\varepsilon$ is the energy of the unperturbed energy levels with zero at $\phi=\pi$, and $E^1_0$ and $E^1_1$ are the perturbed energy levels of the ground and excited states respectively.}
\label{fig:2level}
\end{figure}

The solid lines were calculated using quasi-degenerate perturbation theory~\cite{bransden_00} and show an anti-crossing in the energy levels at $\phi=\pi$. In the system there are many flow states, but we will only consider the lowest two states that represent zero flow and one quantum of clockwise flow. We can ignore the other states if the coupling to them is small compared to their separation in energy. This is the case at $\phi=\pi$, where the ground state, to a good approximation, has the following form,
\begin{equation}
\Psi = a_{0} \psi_0(x) + a_{1} \psi_1(x) + \mbox{corrections}, \label{eq:sup1}
\end{equation}
where $\psi_0$ is an atom with zero flow, $\psi_1$ is an atom with one quantum of clockwise flow, and $a_0$ and $a_1$ are complex amplitudes that satisfy $a_0^2+a_1^2 \approx 1$. In the absence of coupling the energy of the $\psi_0(x)$ and $\psi_1(x)$ states are given by $E_0^0=E^0+\varepsilon(\phi)$ and $E_1^0=E^0-\varepsilon(\phi)$ where $E^0$ is the energy of both states at $\phi=\pi$ and $\varepsilon$ describes the phase dependence of the two energy levels (see Fig.~\ref{fig:2level}). With coupling $V_{01}$ the energies of the ground and first excited state become,
\begin{equation}
E_n = E^0 \pm ( \varepsilon(\phi)^2 + |V_{01}|^2 )^{1/2}, \label{eq:elevel1}
\end{equation}

The amplitudes $a_0$ and $a_1$ are obtained from diagonalising the Hamiltonian and $a_0^2+a_1^2=1$. The ratio of the amplitudes is~\cite{bransden_00},
\begin{equation}
\frac{a_{0}}{a_{1}} = - \frac{V_{01}}{\varepsilon \mp \sqrt{\varepsilon^2 + 
|V_{01}|^2}}. \label{eq:a1overa2}
\end{equation}
Sufficiently near $\phi = \pi$ we see $\varepsilon \ll |V_{01}|$, so $a_{0}/a_{1}\approx 1$ and we get a good superposition as long as $a_0^2+a_1^2 \approx 1$. $\varepsilon$ increases as we move away from $\phi = \pi$, so the zero flow or clockwise flow state's amplitude begins to dominate depending on whether the phase is increased or decreased. To make the superposition more stable we could increase the coupling between the single flow states, but this will increase the coupling to other states and $a_0^2+a_1^2$ will no longer close to $1$. The system can no longer be approximated to a 2-level system, and the simple form of the superposition will be lost.

\begin{figure}
\psfrag{energy}{Energy$/C$} \psfrag{phase}{Phase $\phi$} \psfrag{pi}{$\pi$} \psfrag{pi2}{$\pi/2$} \psfrag{3pi2}{$3\pi/2$} \psfrag{2pi}{$2\pi$}
\includegraphics[width=8cm]{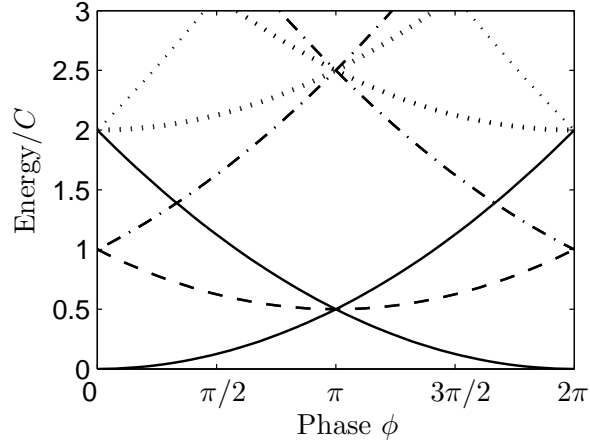}
\caption{\small Show how the energy levels are affected by the phase around the loop. $C = \frac{\hbar^2}{2 m} \left( \frac{2 \pi}{L}  \right) ^2$ is a constant. Some of the lines are now entirely degenerate with other lines.}
\label{fig:spectrum2p}
\end{figure}


For two atoms, the full system, excluding spin, is described by the following Schr\"{o}dinger equation,
\begin{eqnarray}
E_{k_1,k_2}(\phi) \psi_{k_1 k_2}(x_1,x_2) &=& \Big[\frac{(p_{x_1} - \hbar \phi / L)^2}{2 m} + \frac{(p_{x_2} - \hbar \phi / L)^2}{2 m} \nonumber \\
&&+ V_I(x_1 - x_2) + V(x_1) +  V(x_2) \Big] \psi_{k_1 k_2}(x_1,x_2),
\end{eqnarray}
where $V_I(x_1 - x_2)$ is the interatomic potential and $V(x_i)$ is due to a barrier. When $V_I(x_1 - x_2) = 0$ and $V(x_i) = 0$, the wave function is separable and has the form,
\begin{equation}
\psi_{k_1 k_2}(x_1,x_2) = \aleph e^{i 2 \pi k_1 x_1 / L} e^{i 2 \pi k_2 x_2 / L}, \label{eq:soln2p}
\end{equation}
with the associated energies,
\begin{equation}
 E_{k_1,k_2}(\phi) = \frac{\hbar^2}{2 m} \left( \frac{2 \pi}{L} \right) ^2 \left[(k_1+\frac{\phi}{2 \pi})^2 + (k_2+\frac{\phi}{2 \pi})^2 \right],
\end{equation}
where $\aleph = 1/L$. The energy spectrum is shown in Fig.~\ref{fig:spectrum2p}. For $N$ atoms with the same flow, $k$, the energy is simply,
\begin{equation}
E_{k}= \frac{\hbar^2}{2 m} \left( \frac{2 \pi}{L} \right) ^2 N\left(k+\frac{\phi}{2 \pi}\right)^2.
\end{equation}
This shows the gradient of the energy is proportional to the number of atoms in the system, $N$, and thus $\varepsilon$ in Eq.~\ref{eq:a1overa2}, is also proportional to $N$. This means that as the number of particles increases, $a_0/a_1$ will diverge from $1$ more quickly as the phase is changed. For larger systems the range of values of $\phi$ that still produce a cat like state will be narrow, which means it could become experimentally infeasible to produce a superposition. 

For $N >1$ there is no longer a simple crossing of two energy levels at $\phi=\pi$. In the two atom case, the first degeneracy is between four flow states as shown in Fig.~\ref{fig:spectrum2p}. The solid lines are the single flow state $\psi_{00}(x_1,x_2)$ and $\psi_{11}(x_1,x_2)$, while the dashed line is both the multiple flow states $(\psi_{01}(x_1,x_2) \pm \psi_{01}(x_2,x_1))/\sqrt{2}$, which are degenerate for all $\phi$. To create a superposition we need to increase the energy of the multiple flow states above the single flow states. This will reduce the amplitude of the multiple flow states in the ground state of the system. To do this we consider the interaction energy between the particles.

To see how the desired separation might arise we assume the interaction energy is positive and it is the same between all particles. A system consisting of 2 identical bosons with different flow has wave function,
\begin{equation}
\psi_S (k_1, k_2) = \frac{1}{\sqrt{2}} \left[ \psi(k_1, k_2) + \psi(k_2, k_1)\right],
\end{equation}
where $\psi(k_1, k_2)=\psi_{k_1}(x_1)\psi_{k_2}(x_2)$ and $\psi(k_2, k_1)=\psi_{k_2}(x_1)\psi_{k_1}(x_2)$. If the particles are in the same flow state then the wave function is simply $\psi(k, k)$. The expectation value for the interaction energy can be calculated using the interatomic potential $V(|x_1 - x_2|)$. This gives,
\begin{equation}
\langle \psi (k, k) | V(|x_1 - x_2|) | \psi (k, k) \rangle = V,
\end{equation}
for all atoms in the same flow state. For atoms in different flow states we get,
\begin{eqnarray}
\langle \psi_S (k_1, k_2)\!\!\! &&\!\!\!| V(|x_1 - x_2|) | \psi_S (k_1, k_2) \rangle = \frac{1}{2} [ \langle  \psi(k_1, k_2) | V(|x_1 - x_2|) | \psi(k_1, k_2) \rangle \nonumber \\
&&+ \langle  \psi(k_2, k_1) | V(|x_1 - x_2|) | \psi(k_2, k_1) \rangle + \langle  \psi(k_1, k_2) | V(|x_1 - x_2|) | \psi(k_2, k_1) \rangle \nonumber \\
&&+ \langle  \psi(k_2, k_1) | V(|x_1 - x_2|) | \psi(k_1, k_2) \rangle ] = V + (V_{12,21} + V_{21,12})/2,
\end{eqnarray}
where $V = \langle \psi(k_i, k_j) | V(|x_1 - x_2|) | \psi(k_i, k_j)\rangle$ and $V_{ij,kl} = \langle  \psi(k_i, k_j) | V(|x_1 - x_2|) | \psi(k_k, k_l) \rangle$. The interaction energy has more terms when the atoms have different flow. If the interaction term is a short range $\delta$-function, the energy shift is $2V$ from the non-interacting case. We shall consider this case in more detail when we look at the Bose-Hubbard model. If we could use atoms with ``medium'' range interactions, e.g. dipolar ones~\cite{goral_02}, then there would be an additional flow state dependence. This is explored in section~\ref{sec:LRF}.

As the number of atoms is increases more atoms can have different flow, which increases the energy shift from the non-interacting case. For three atoms the maximum energy shift is $6V$ from the non-interacting case, which happens when all atoms have different flow. If some of the atoms have the same flow the energy difference is smaller and the coupling to the ground state is stronger. To produce a good cat state we need to separate the single flow states from the excited states. This is essentially what is realised in successful experiments to date. They achieve this in various ways. In the experiment described in Ref.~\cite{arndt_99} it is the bound nature of the C${}_{60}$ molecule that separates excited states from the two close lying states i.e. the molecule moving in two different directions. In the case of the SQUID the plasma frequency gives a gap to the excited states that have Cooper pairs with different flow~\cite{friedman_00}. The coupling to other flow states, which reduces the ability to make cat states, is similar to decoherence where the macroscopic state is coupled to other states via interactions with the environment. There is however a crucial difference, because the coupling is due to the system itself and can be reduced by making the barrier smaller, although this requires more accurate control of macroscopic parameters to make a cat state.

From Eq.~\ref{eq:a1overa2} we see we need $|V_{01}| \gg \varepsilon$ for $a_0/a_1 \approx 1$ along with $a_0^2+a_1^2\approx 1$. $a_0^2+a_1^2$ decreases as the coupling to other states increases, because other flow states will have a significant amplitude. For $N$ atoms with short range s-wave scattering the interaction potential has the form,
\begin{equation}
V_I = V \sum_{i\neq j}^N \delta(x_i-x_j).
\end{equation}
The terms that affect the splitting are then given by,
\begin{eqnarray}
V_{00}-V_{11} &=& \langle \psi_0 | V_I | \psi_0 \rangle = 0, \\
|V_{01}| &=& \frac{ V}{L^N} \frac{N}{2} (N-1) \int_0^L dx_1...dx_N e^{2 \pi (x_1+...+x_N)/L}\delta(x_1-x_2) \nonumber \\
&=& V\frac{N}{2} \frac{(N-1)}{2L} \frac{1}{(2 \pi)^{N-1}},
\end{eqnarray}
where, $\psi_0 = 1 / L^{N/2}$ and $\psi_1 = e^{i 2 \pi x_1 /L} e^{i 2 \pi x_2 /L}... e^{i 2 \pi x_N /L} /L^{N/2}$. As the number of particles increases the $(2\pi)^{N-1}$ term increases exponentially while the $N(N-1)/2$ term increases quadratically, so the overall coupling term, $|V_{01}|$, gets smaller. This makes it harder to get $a_0/a_1 \approx 1$, therefore cat states become harder to make.

The physical basis behind the effective coupling is the $N(N-1)/2$ ways of coupling from $|N,0\rangle$ to $|0,N\rangle$. This coupling is, of course, effected directly by the energy defect to intermediate states. This makes the transition less likely for larger systems, because more energy is needed for all atoms to make the transition and gives the $(2\pi)^{N-1}$ term, which decreases the energy gap at the anti-crossing. For example, to get from $\psi(0,0)$ to $\psi(1,1)$ there is coupling through $\psi(0,1)$, while the shortest path to couple from $\psi(0,0,0)$ to $\psi(1,1,1)$ is through $\psi(0,0,1)$ then $\psi(0,1,1)$. In the next section this will be investigated further with a more realistic system, the Bose-Hubbard model.

\section{Coupled BECs in a ring}
\label{sec:BEC}

\begin{figure}[t]
\psfrag{phi}{$\phi/3$} \psfrag{J1}{$J_1$} \psfrag{J2}{$J_2$} 
\psfrag{J3}{$J_3$} \psfrag{a}{\LARGE $a$} \psfrag{c}{\LARGE $b$} 
\psfrag{b}{\LARGE $c$}
\includegraphics[width=6cm]{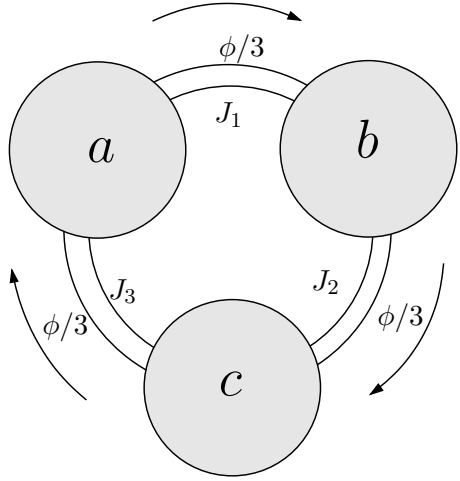}
\caption{\small The system consists of three sites, $a$, $b$ and $c$, where atoms can tunnel between sites with tunnelling strengths $J_1$, $J_2$ and $J_3$ and there is a clockwise phase difference of $\phi/3$ between each site.}
\label{fig:setup}
\end{figure}

Hallwood \emph{et al.}~\cite{hallwood_06} presented a scheme for producing a multiparticle superposition of different superfluid flow states in a ring of coupled BECs using the BHM. In this scheme, it was possible to simulate the production of superpositions of all atoms stationary and all atoms rotating clockwise by applying a $\pi$ phase around a loop and changing the tunnelling strength between the sites. This was done using an adapted Bose-Hubbard model~\cite{rey_03} that allows a change in applied phase between three sites that form a ring (see Fig.~\ref{fig:setup}). This gives the ``twisted'' Hamiltonian,
\begin{equation}
H = -J[ e^{i\phi/3}\left( a^\dagger b + b^\dagger c + c^\dagger a\right) 
+ e^{-i\phi/3}\left(b^\dagger a +  
c^\dagger b + a^\dagger c \right)] + U ( {a^\dagger}^2 a^2 + {b^\dagger}^2 b^2 + {c^\dagger}^2 c^2),
\label{eq:ham_num}
\end{equation}
where $a$, $b$ and $c$ are the annihilation operators of atoms on the three sites, $U$ is the on-site interatomic interaction strength and $J$ is the tunnelling strength between adjacent sites. The phase factors $e^{\pm i \phi/3}$ in the coupling terms are known as Peierls phase factors~\cite{rey_03}. We note that $\phi$ does not have to obey the phase matching condition because it represents applied phase and not the phase of the condensate.

Using this scheme we want to understand how the ground state is effected by experimental parameters and how cat states can be formed. The parameters we consider are the applied phase, $\phi$, which needs to be $\sim\pi$, the interaction strength, $U$, the coupling strength, $J$, and the number of atoms, $N$. The value of $U/J$ determines the quantum phase of the system, giving a superfluid for $U/J\ll1$ and a insulator for $U/J\gg1$. We need to be in the near superfluid regime to achieve a good cat state, so atoms can tunnel freely, while there is still coupling between flow states. The state of the system can be found by direct diagonalisation, but there is a limit to how many atoms can be used of order $40$ due to computational time. It is therefore necessary to find an approach that can look at higher numbers, which can be checked by comparison with exact calculations for lower numbers.

It is convenient to consider a new complete orthogonal basis of operators that incorporates the phase matching condition, $\{ \alpha, \beta, \gamma \}$. We will use the quasi-momentum basis, or flow basis, because we will be producing superpositions of single flow states,   
\begin{eqnarray}
&&\alpha = (a + b + c) / \sqrt{3},\nonumber\\ 
&&\beta = (a + b e^{i 2 \pi / 3} + c e^{i 4 \pi / 3}) / \sqrt{3},\nonumber \\
&&\gamma = (a + b e^{- i 2 \pi / 3} + c e^{- i 4 \pi / 3}) / \sqrt{3}.
\label{eq:quasimodes}
\end{eqnarray}
The new operator basis corresponds respectively to annihilation of an atom with zero flow, one quantum of clockwise flow and one quantum of anticlockwise flow. They obey the usual commutation relations and the quasi-momentum conservation rules for a periodic potential. Using these operators we can rewrite the ``twisted'' Hamiltonian in flow representation,
\begin{eqnarray}
H &=& - J \{ (2 \alpha^{\dagger} \alpha - \beta^{\dagger} \beta - \gamma^{\dagger} \gamma) \cos(\phi/3) + \sqrt{3}(\beta^{\dagger} \beta - \gamma^{\dagger} \gamma) \sin(\phi/3) \} \nonumber \\
&&+ \frac{U}{3} \{ \alpha^{\dagger}{}^2 \alpha^2 + \beta^{\dagger}{}^2 \beta^2 + \gamma^{\dagger}{}^2 \gamma^2 + 4(\alpha^{\dagger} \alpha \beta^{\dagger} \beta + \alpha^{\dagger} \alpha \gamma^{\dagger} \gamma  + \beta^{\dagger} \beta\gamma^{\dagger} \gamma ) \nonumber \\
&&+ 2( \alpha^2 \beta^{\dagger} \gamma^{\dagger} + \beta^2 \alpha^{\dagger} \gamma^{\dagger} + \gamma^2 \alpha^{\dagger} \beta^{\dagger} + h.c. \}. \label{eq:ham_mom}
\end{eqnarray}
From Eq.~(\ref{eq:ham_mom}) we see that the eigenstates of the system when $U/J \! \ll \! 1$ are just flow states. For $U/J \! \approx \! 1$, there is coupling between the different flow states and when $U/J \! \gg \! 1$ the system is in the Mott regime where each site acquires the same number of atoms. 

We are interested in finding a superposition of the form,
\begin{equation}
|\Psi\rangle = a_0 | N,0,0 \rangle + a_1 | 0,N,0 \rangle, \label{eq:sup3}
\end{equation}
where $| N,0,0 \rangle$ is the state where all atoms are stationary and $| 0,N,0 \rangle$ is the state where all atoms have one unit of clockwise flow. The terms in the ket represent the number of atoms in the $\alpha$ , $\beta$ and $\gamma$ modes respectively. To study how good the produced cat state is, we calculate the ratio of the two amplitudes $a_0/a_1$ as we did in section~\ref{sec:loop}. We consider systems where $a_0^2+a_1^2\approx1$, so a good cat state is formed when $a_0/a_1 \approx 1$. For $a_0^2+a_1^2\approx1$ we require the system to be in the near superfluid regime, but we will see that this has a detrimental effect on the ratio $a_0/a_1 \approx 1$ when $\phi$ is not exactly $\pi$. We require $\phi=\pi$ for the energy of the flow states $| N,0,0 \rangle$ and $| 0,N,0 \rangle$ to be degenerate. It is, of course, important to see how small deviations from a $\pi$ phase change will effect the quality of the superposition. This will enable us to assess the precision we shall need in an experiment.

For large numbers of atoms the system states can be extremely complex. For now we will only consider atoms with all the same flow. In certain regimes this approximation is justified, because the atom-atom interactions raises the energy of multiple flow states faster than single flow states, so $a_0^2+a_1^2\approx1$.

From Fig.~\ref{fig:2level} we see the flow state energy levels are given by $E_0^0 = E^0-\varepsilon$ and $E_1^0 = E^0+\varepsilon$. Using Eq.~\ref{eq:sup3} we can reduce the problem to a $2 \times 2$ matrix Hamiltonian as in section~\ref{sec:loop}. The energy and the amplitudes are described by Eq.~\ref{eq:elevel1} and Eq.~\ref{eq:a1overa2}. This shows the ratio of the amplitudes only depends on the coupling between the two flow states, $V_{01}$, and the rate at which the energy levels change, $\varepsilon$. For a stable superposition we would require $V_{01} \gg \varepsilon$, but for large $V_{01}$ the perturbation must be big, so other states will begin to have a significant amplitude in the ground state.

To use Eq.~\ref{eq:a1overa2} we need to understand how $\varepsilon$ and $\phi$ are related. The Hamiltonian for the case of no atom-atom interaction, i.e. $U=0$, is diagonal in the flow basis,
\begin{equation}
H_0 = - J \{ (2 \alpha^{\dagger} \alpha - \beta^{\dagger} \beta - \gamma^{\dagger} \gamma) \cos(\phi/3) - \sqrt{3} (- \beta^{\dagger} \beta + \gamma^{\dagger} \gamma) \sin(\phi/3) \}. \label{eq:ham0}
\end{equation}
The change in the Hamiltonian due to interactions is then given by,
\begin{eqnarray}
H' &=& \frac{U}{3} \{ \alpha^{\dagger}{}^2 \alpha^2 + \beta^{\dagger}{}^2 \beta^2 + \gamma^{\dagger}{}^2 \gamma^2 + 4(\alpha^{\dagger} \alpha \beta^{\dagger} \beta + \alpha^{\dagger} \alpha \gamma^{\dagger} \gamma + \beta^{\dagger} \beta\gamma^{\dagger} \gamma ) \nonumber \\
&&+ 2( \alpha^2 \beta^{\dagger} \gamma^{\dagger} + \beta^2 \alpha^{\dagger} \gamma^{\dagger} + \gamma^2 \alpha^{\dagger} \beta^{\dagger} + h.c. \}.
\label{eq:ham'}
\end{eqnarray}
From this we can see that the unperturbed ground state energy has the form $E_1 = -2JN \cos(\phi/3)$. If we subtract the energy of the two states at the crossing, $\phi = \pi/3$, we get,
\begin{equation}
\varepsilon = E_0^0(\pi)- 2JN \cos(\phi/3)=JN -2JN \cos(\phi/3). \label{eq:epsilon}
\end{equation}
This shows that the gradient of the energy increases linearly with the number of atoms, and thus, the ratio ${a_0 / a_1}$ will diverge from $1$ more quickly when $\phi$ moves away from $\pi$ as more particles are added. This will, of course, make it more and more difficult to produce a cat state.

\begin{figure}[ht]
\psfrag{300}{$|3,0,0\rangle$} \psfrag{210}{$|2,1,0\rangle$} \psfrag{120}{$|1,2,0\rangle$} \psfrag{030}{$|0,3,0\rangle$} 
\psfrag{201}{$|2,0,1\rangle$} \psfrag{111}{$|1,1,1\rangle$} \psfrag{021}{$|0,2,1\rangle$} \psfrag{102}{$|1,0,2\rangle$} 
\psfrag{003}{$|0,0,3\rangle$} \psfrag{012}{$|0,1,2\rangle$}
\psfrag{400}{$|4,0,0\rangle$} \psfrag{310}{$|3,1,0\rangle$} \psfrag{220}{$|2,2,0\rangle$} \psfrag{130}{$|1,3,0\rangle$}
\psfrag{040}{$|0,4,0\rangle$}
\psfrag{301}{$|3,0,1\rangle$} \psfrag{211}{$|2,1,1\rangle$} \psfrag{121}{$|1,2,1\rangle$} \psfrag{031}{$|0,3,1\rangle$} 
\psfrag{202}{$|2,0,2\rangle$} \psfrag{112}{$|1,1,2\rangle$} \psfrag{022}{$|0,2,2\rangle$} 
\psfrag{103}{$|1,0,3\rangle$} \psfrag{013}{$|0,1,3\rangle$}
\psfrag{004}{$|0,0,4\rangle$}
\includegraphics[width=7cm]{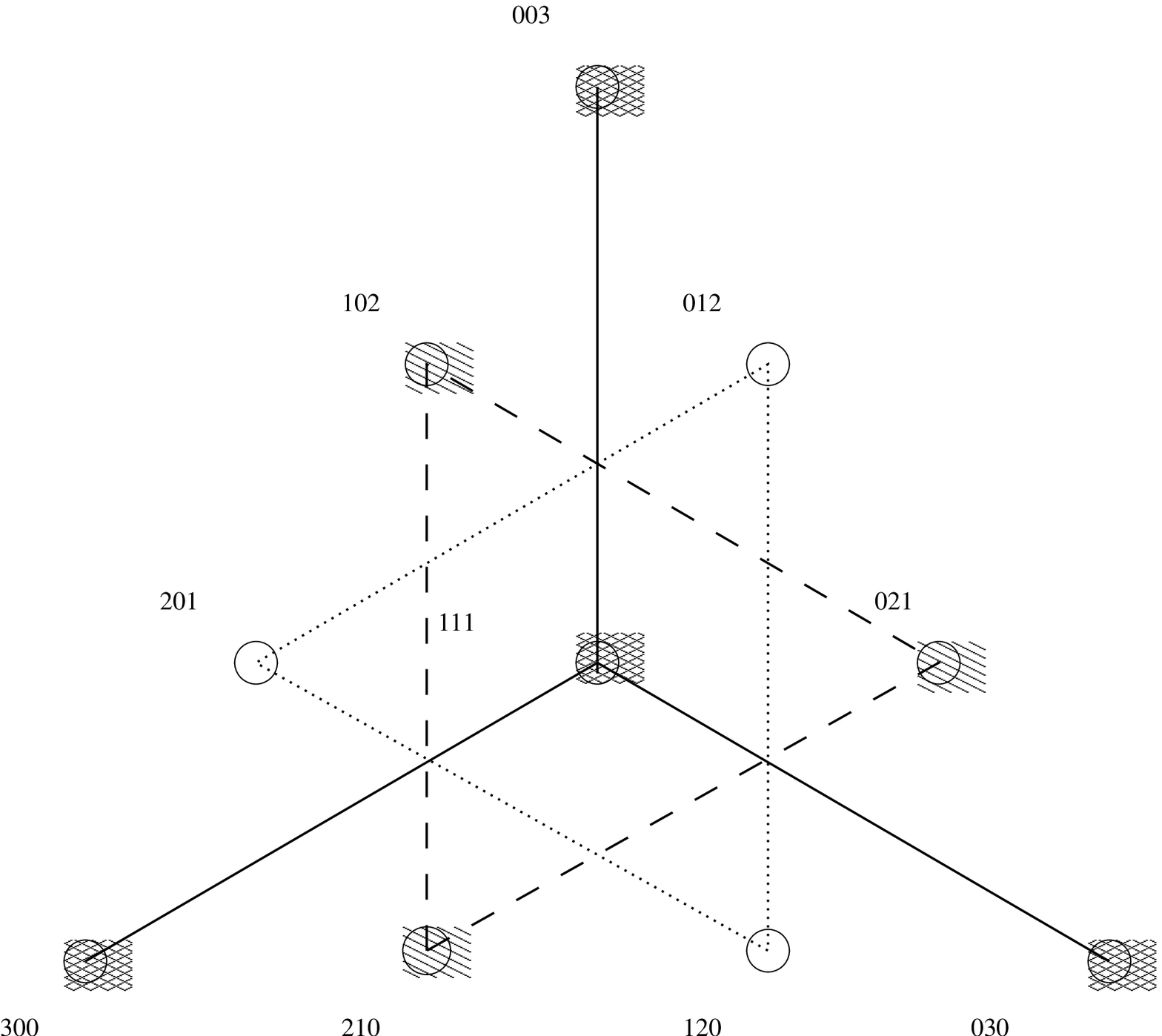}
\includegraphics[width=7cm]{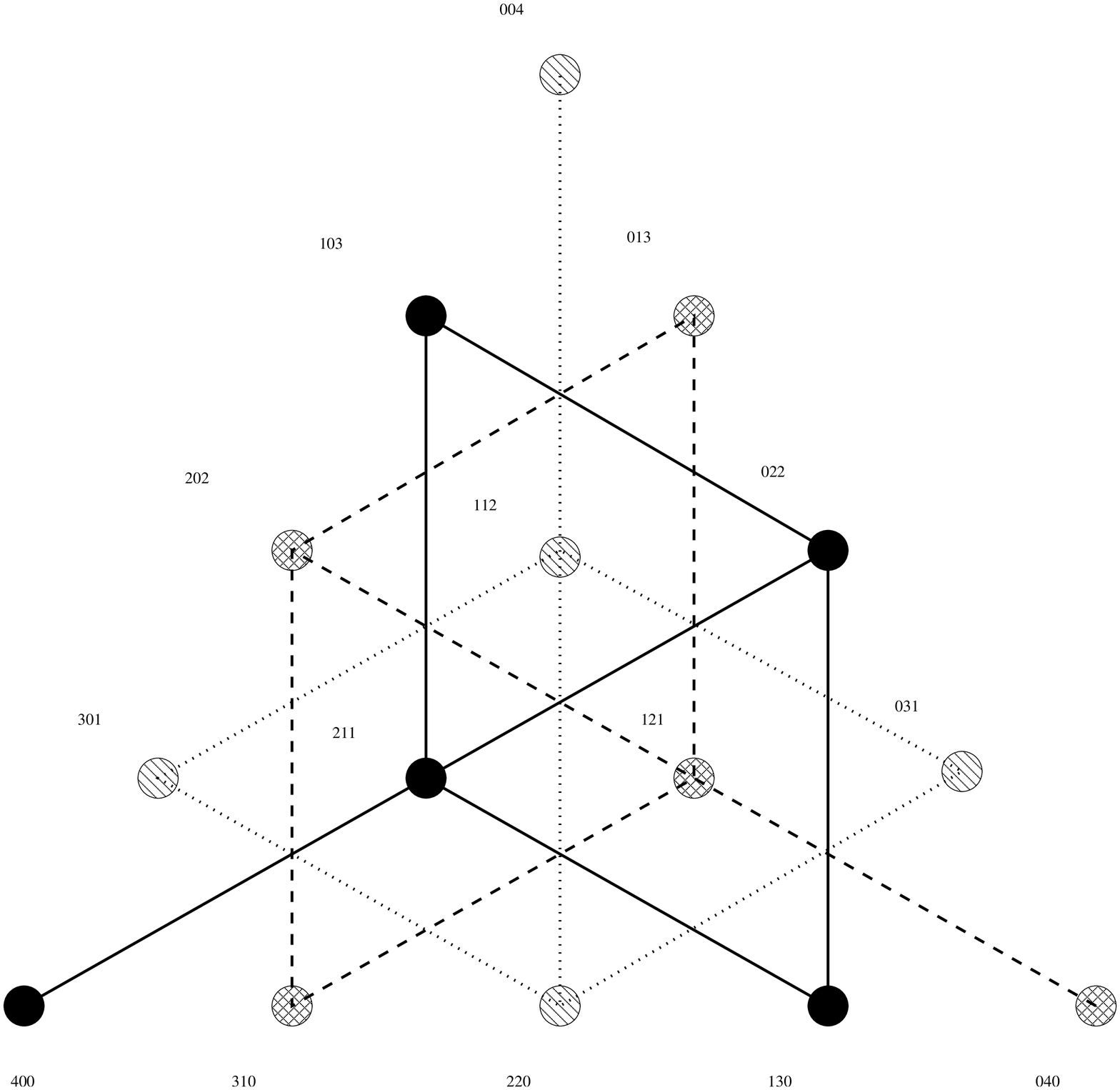}
\caption{\small Left: All the possible flow states for 3 atoms. The different colours show the coupling between different state. Right: All the possible flow states for 4 atoms. It is no longer possible to connect the pure flow states, so no superposition is possible between these states.}
\label{fig:coupling_atoms}
\end{figure}

\begin{figure}[ht]
\psfrag{400}{$|4,0,0\rangle$} \psfrag{310}{$|3,1,0\rangle$} \psfrag{220}{$|2,2,0\rangle$} \psfrag{130}{$|1,3,0\rangle$}
\psfrag{040}{$|0,4,0\rangle$}
\psfrag{301}{$|3,0,1\rangle$} \psfrag{211}{$|2,1,1\rangle$} \psfrag{121}{$|1,2,1\rangle$} \psfrag{031}{$|0,3,1\rangle$} 
\psfrag{202}{$|2,0,2\rangle$} \psfrag{112}{$|1,1,2\rangle$} \psfrag{022}{$|0,2,2\rangle$} 
\psfrag{103}{$|1,0,3\rangle$} \psfrag{013}{$|0,1,3\rangle$}
\psfrag{004}{$|0,0,4\rangle$}
\includegraphics[width=7cm]{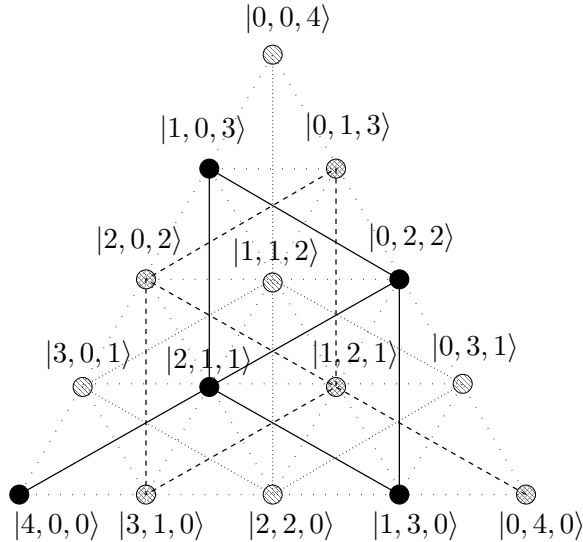}
\caption{\small All the possible flow states for 4 atoms. The Hamiltonian that represents the coupling in this diagram has unequal tunnelling strengths, so coupling between the pure flow states is possible.}
\label{fig:coupling_atomsJ}
\end{figure}

From Eq.~\ref{eq:ham_mom} we see there is no direct coupling from $| N,0,0 \rangle$ to $|0,N,0 \rangle$, so we must consider other flow states to calculate the total coupling through intermediate state.
 Note when $U=0$ there are no off diagonal terms and the coupling is not possible. In this case the eigenfunctions are just flow states. To understand what we mean by coupling path and intermediate states, lets consider a 4 state system where each state is represented by one of $|1 \rangle$, $|2 \rangle$, $|3 \rangle$ and $|4 \rangle$ with Hamiltonian,
\begin{equation}
H= \left( \begin{array}{cccc}
		\varepsilon_1 & V_{12} & V_{13} & V_{14} \\
		V_{21} & \varepsilon_2 & V_{23} & V_{24} \\
		V_{31} & V_{32} & \varepsilon_3 & V_{34} \\
		V_{41} & V_{42} &  V_{43} & \varepsilon_4
       \end{array} \right),
\end{equation}
where $\varepsilon_i$ is the uncoupled energy of state $i$ and $V_{jk}$ is the coupling from state $j$ to $k$. Although there is direct coupling from $| 1 \rangle$ to $| 4 \rangle$ there are also indirect couplings through intermediate states $| 2 \rangle$ and $| 3 \rangle$. This is made more explicit in the four coupled equations that constitute the Schr\"{o}dinger equation of the system. By eliminating the intermediate states we can simplify the problem to a two level system and a $2\times 2$ matrix Hamiltonian. This gives a combined coupling from $|1 \rangle$ to $|4 \rangle$ of,
\begin{eqnarray}
V_{\mbox{\small comb}} &=& V_{14} - \frac{V_{14}V_{23}V_{32}}{(\lambda -
\varepsilon_2)(\lambda - \varepsilon_3)}
                + \frac{V_{12}V_{24}}{(\lambda - \varepsilon_2)}
		+ \frac{V_{13}V_{34}}{(\lambda - \varepsilon_3)} \nonumber \\
		&&+ \frac{V_{12}V_{23}V_{34}}{(\lambda-\varepsilon_2)(\lambda-\varepsilon_3)}
		+ \frac{V_{13}V_{32}V_{24}}{(\lambda-\varepsilon_2)(\lambda-\varepsilon_3)}, \label{eq:coupling}
\end{eqnarray}
where $\lambda$ is the ground state energy of the Hamiltonian that can be calculated using perturbation theory. If $\varepsilon_2$ and $\varepsilon_3 \gg \varepsilon_1$ and $\varepsilon_4$, and $\varepsilon_1 \approx \varepsilon_4$ then the ground state will have most of its amplitude in states $|1\rangle$ and $|4\rangle$, and a superposition between $|1\rangle$ and $|4\rangle$ is formed. What this shows us is the coupling is the addition of all the different paths from state $|1 \rangle$ to $|4\rangle$ where the individual terms are given by the coupling to the intermediate states divided by $(\lambda-\varepsilon_i)$.

In the Bose-Hubbard model we can define a complete basis that describes all the different numbers of atoms with each flow where the total number of atoms is $N$. The coupling between these states is more clearly shown if we draw all the possible states in flow space and draw lines to show the possible coupling. This has been done for 3 and 4 atoms shown in Fig.~\ref{fig:coupling_atoms} (a) and (b). In Fig.~\ref{fig:coupling_atoms} (a) there is a coupling from $|3,0,0\rangle$ to $|0,3,0\rangle$ and $|0,0,3\rangle$ through $|1,1,1\rangle$, so under the correct conditions it will be possible to form a superposition between any two of the three single flow states. This happens when the phase is half way between the phase of the two superposing states, so the energy of the two states is degenerate. In Fig.~\ref{fig:coupling_atoms} (b) there is no path between the $|4,0,0\rangle$, $|0,4,0\rangle$ and $|0,0,4\rangle$ states, so there is no coupling and, as has been shown~\cite{hallwood_06}, no superposition is formed. This idea can be applied to higher numbers of atoms and it was found that coupling is only possible for commensurate numbers of atoms, i.e. the number of atoms divided by the number of sites must be an integer.

This problem is resolved by adding different tunnelling strengths between the three sites, so the Hamiltonian now has terms coupling all adjacent flow states and Fig.~\ref{fig:coupling_atoms} (b) becomes Fig.~\ref{fig:coupling_atomsJ}. This shows there are many different paths from one pure flow state to the other although the strength of the coupling depends on the differential tunnelling strength and atom-atom interaction strength.

We were initially interested in seeing how this coupling changes as the number of atoms increases. Eq.~\ref{eq:coupling} gave the form of the coupling from state $|1\rangle$, to $|4\rangle$ via states $|2\rangle$ and $|3\rangle$. Generalising to larger numbers of intermediate states we get,
\begin{equation}
V_{1n}= \frac{V_{1i} V_{ij}...V_{pn}}{(\lambda -\varepsilon_i)(\lambda -\varepsilon_j)...(\lambda -\varepsilon_p)} + \mbox{other possible paths}.
\end{equation}
All the different paths must be added together and each path has a combined coupling term, which is the product of the individual coupling terms along that path divided by $(\lambda -\varepsilon_j)$ of each of the intermediate states, where $\varepsilon_j$ is the diagonal matrix element in the Hamiltonian of that intermediate state. As the number of atoms increases the length of the combined coupling increases and because $V_{ij}$ is generally smaller than $(\lambda -\varepsilon_j)$, so the overall coupling will decrease, even though the number of paths increases. From Eq.~\ref{eq:a1overa2} we see that the ratio $a_0/a_1$ will move away from $1$ more rapidly for larger numbers of atoms as $\phi$ deviates from $\pi$. This will make it correspondingly harder to make a clean cat state.

\begin{figure}[ht]
\psfrag{phi}{$\Delta \phi$} \psfrag{a1overa2}{$a_1/a_2$} \psfrag{a}{(a)}\psfrag{b}{(b)}  \psfrag{c}{(c)} \psfrag{d}{(d)} \psfrag{N3}{$N=3$} \psfrag{N6}{$N=6$} \psfrag{N9}{$N=9$} \psfrag{N12}{$N=12$}
\psfrag{simulated}{Simulated line}
\includegraphics[width=6cm]{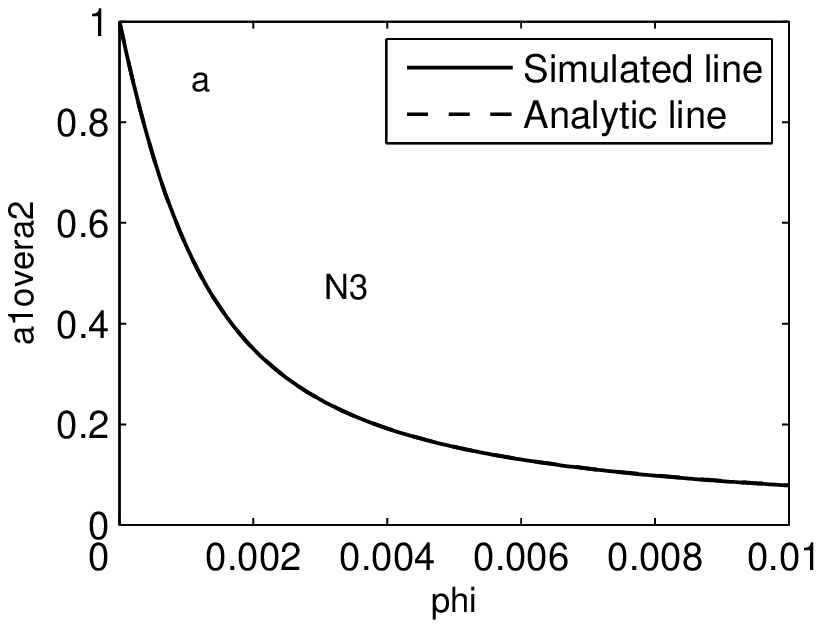}
\includegraphics[width=6cm]{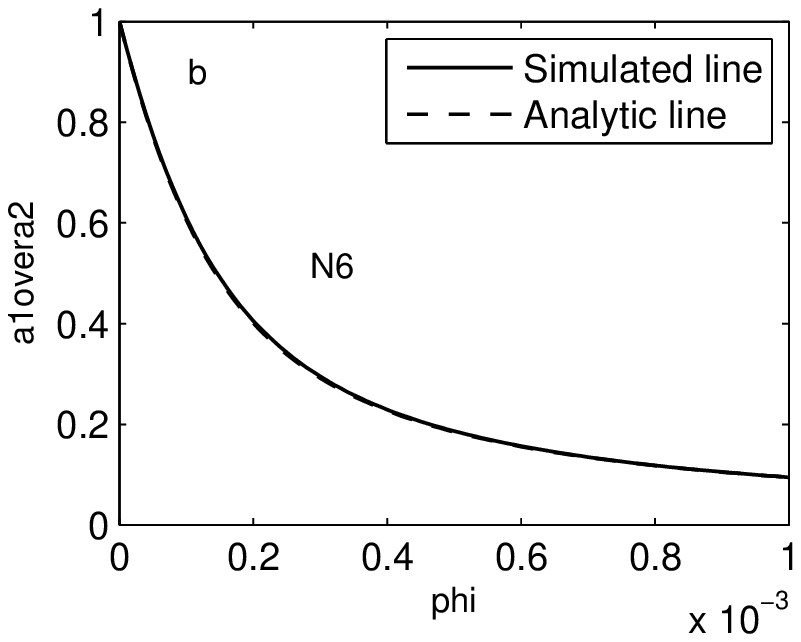}
\includegraphics[width=6cm]{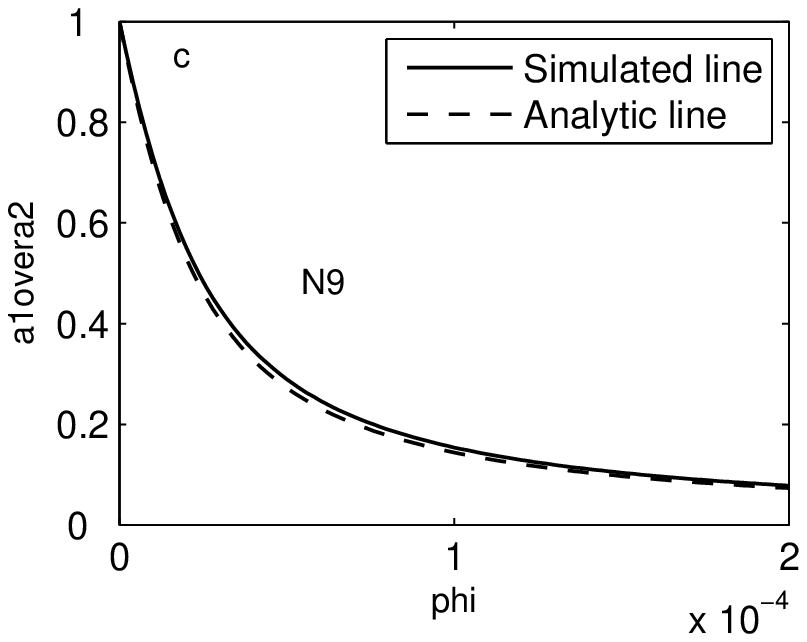}
\includegraphics[width=6cm]{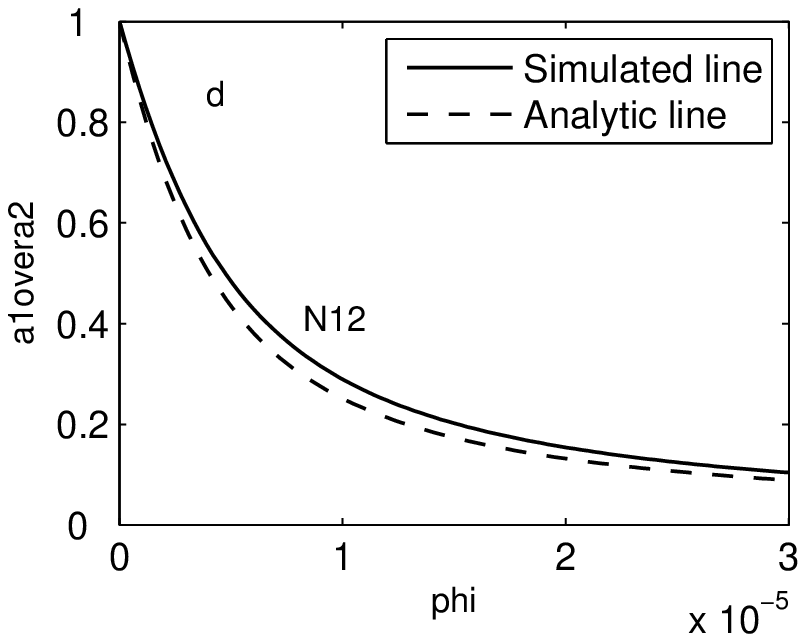}
\caption{\small Simulated and analytic results for the effect of changing the phase by $\Delta \phi$ away from $\pi/3$ for (a) 3 atoms, (b) 6 atoms, (c) 9 atoms and (d) 12 atoms, where $U/J=0.1$.}
\label{fig:phaseVcattiness1}
\end{figure}

The combined coupling between the two single flow states has been calculated for $N=3$, $6$, $9$ and $12$ atoms and using Eq.~\ref{eq:a1overa2} and~\ref{eq:epsilon} we have calculated $a_0/a_1$ as a function of $\Delta \phi=\phi-\pi$ as shown in Fig.~\ref{fig:phaseVcattiness1}. This can also be done for non commensurate numbers with unequal $J$, where there are many more coupling paths. Fig.~\ref{fig:phaseVcattiness1} show the simulated and analytic results for the effect on $a_0 / a_1$ when the phase is not exactly $\pi$ around the loop. The analytic result for low numbers fit the simulated results very well. As the number of atoms increase the quality of the fit is reduced. This is due to the large number of states, so the total amplitude in the  $|N,0,0 \rangle$ and $|0,N,0 \rangle$ states is reduced. One can see in Fig.~\ref{fig:phaseVcattiness1} the slopes drops more quickly as the number of atoms in the system is increased. This shows how experimentally difficult it will be to create a cat state, as the number of atoms is increased.

\section{Modifications to the model}
\label{sec:LRF}
The discussion in Sections~\ref{sec:loop} and~\ref{sec:BEC} have demonstrated clearly that as the number of atoms is raised it becomes increasingly difficult to make a superposition. The analysis suggests that even for relatively moderate numbers it would be experimentally impossible to make superpositions. We know, however, that SQUID experiments~\cite{friedman_00, wal_00} have shown it is possible, so what feature is missing from our model? We believe that the crucial difference is our use of short-ranged interactions. In this section we shall, therefore consider the effect of long range interactions. We shall also consider briefly how having a continuous loop with a barrier, and not three sites, will improve our chances of making a cat state. 

We shall suppose that the long range interaction is due to either magnetic or electric dipole-dipole interactions between the atoms. An electric dipole-dipole interaction can be induced by applying a strong d.c. electric field across the condensate~\cite{marinescu}. It has also been shown to be possible to trap atoms with larger magnetic moments, so interactions beyond s-waved scattering must be considered~\cite{kim}. There has in fact been a good deal of experimental and theoretical work on the dipole-dipole interaction~\cite{goral_00,santos_00,goral_02a,damski_03,afrousheh_04}. In this paper we shall use the formalism presented in the paper by G\/{o}ral \emph{et al.}~\cite{goral_02}, where the magnetic dipole-dipole interaction potential is thus given by~\cite{goral_00},
\begin{equation}
V_{dd} = \frac{\mu_0}{4\pi} \frac{{\bf\mu}_1({\bf r})\cdot{\bf \mu}_2({\bf r}) 
          - 3({\bf\mu}_1({\bf r})\cdot{\bf u})({\bf \mu}_2({\bf r})\cdot{\bf u})}{|{\bf r}-{\bf r}'|^3},
\end{equation}
where ${\bf u}$ is the vector joining the interacting particles, $\mu_0$ is the magnetic permeability of the vacuum and $\mu_i$ are the magnetic moments of the atoms. This can be simplified by assuming the magnetic moments of each atom points in the same direction. Adding the s-wave scattering length potential, the total interaction potential then takes the form~\cite{goral_02},
\begin{equation}
V_{int} =  \frac{4\pi\hbar^2a}{m}\delta({\bf r}-{\bf r}') + d^2 \frac{1-3\cos^2\theta}{|{\bf r}-{\bf r}'|^3}.
\end{equation}
The first term is the usual short range interaction with an $s$-wave scattering length of $a$ and atomic mass $m$. The second term is due to the dipole-dipole interaction of strength $d$. As we are only dealing with a 2-D plane we can set $\theta = \pi/2$. The dipole-dipole interaction produces off site interactions in the BHM, so for the three site case there are nearest neighbour interactions only, and the Bose-Hubbard Hamiltonian becomes,
\begin{eqnarray}
H &=& -J[ e^{i\phi/3}\left( a^\dagger b + b^\dagger c + c^\dagger a\right) + e^{-i\phi/3}\left(b^\dagger a + c^\dagger b + a^\dagger c \right)] \nonumber \\
   && + U_0 ( {a^\dagger}^2 a^2 + {b^\dagger}^2 b^2 + {c^\dagger}^2 c^2) + U_1 ( (a^{\dagger})^2 b^2 + (b^{\dagger})^2 c^2 + (c^{\dagger})^2 a^2), \label{eq:ham_dip}
\end{eqnarray}
where $a$, $b$ and $c$ are the usual annihilation operators for the three sites, $J$ is the tunnelling strength and $U_0$ is the on-site interaction strength. The extra term $U_1$ is the strength of the interaction between atoms on the adjacent sites. The interaction terms are given by,
\begin{equation}
U_{\sigma} = \int |w({\bf r}-{\bf r}_i)|^2 V_{int}({\bf r}-{\bf r}')|w({\bf r}'-{\bf r}_j)|^2 d^3r d^3r',
\end{equation}
where $|r_i-r_j|=4\pi \sigma /|{\bf k}|$, and $i$ and $j$ are the indices of the sites. $w({\bf r}-{\bf r}_i)$ are Wannier functions used to describe the field operators in the Hamiltonian and are well localised on the lattice sites~\cite{jaksch_98}. The two interaction terms are therefore given by,
\begin{equation}
U_{0} = \frac{4\pi\hbar^2a}{m} \int |w({\bf r}-{\bf r}_0)|^4 d^3r + d^2 \int \frac{|w({\bf r}-{\bf r}_0)|^2 |w({\bf r}'-{\bf r}_0)|^2}{|{\bf r}-{\bf r}'|^3} d^3r d^3r', \label{eq:on_int}
\end{equation}
and
\begin{equation}
U_{1} = d^2 \int \frac{|w({\bf r}-{\bf r}_0)|^2 |w({\bf r}'-{\bf r}_1)|^2}{|{\bf r}-{\bf r}'|^3} d^3r d^3r'.
\end{equation}
The integrand in the second integral in Eq.~\ref{eq:on_int} becomes infinite for ${\bf r}'={\bf r}$, which is the case for on site interactions. This problem arises due to the pseudo-potential form of the dipole-dipole interaction. A better representation of the potential is required to solve this, which is investigated by Ronen \emph{et al.}~\cite{ronen_06}. Even if the $s$-wave scattering is small and the dipole-dipole interaction is large we can see that $U_1$ is always smaller than $U_0$. 

Again we are interested in the flow of the atoms, rather than position, so we transform to the flow basis defined by Eq.~\ref{eq:quasimodes}. Eq.~\ref{eq:ham_dip} then takes the form:
\begin{eqnarray}
H &=& - J \{ (2 \alpha^{\dagger} \alpha - \beta^{\dagger} \beta - \gamma^{\dagger} \gamma) \cos(\phi/3) + \sqrt{3}(\beta^{\dagger} \beta - \gamma^{\dagger} \gamma) \sin(\phi/3) \} \nonumber \\
&&+ \frac{(U_0+U_1)}{6} \{ \alpha^{\dagger}{}^2 \alpha^2 + \beta^{\dagger}{}^2 \beta^2 + \gamma^{\dagger}{}^2 \gamma^2\} + \frac{(4U_0+U_1)}{6}\{(\alpha^{\dagger} \alpha \beta^{\dagger} \beta + \alpha^{\dagger} \alpha \gamma^{\dagger} \gamma  + \beta^{\dagger} \beta\gamma^{\dagger} \gamma )\} \nonumber \\
&&+ \frac{(2U_0-U_1)}{6}\{( \alpha^2 \beta^{\dagger} \gamma^{\dagger} + \beta^2 \alpha^{\dagger} \gamma^{\dagger} + \gamma^2 \alpha^{\dagger} \beta^{\dagger} + h.c. \}. \label{eq:ham_mom_d}
\end{eqnarray}
In order to get a good cat we need the extra long range interactions to increase the difference in energy between the single flow states and the multiple flow states. Examining the terms in Eq.~\ref{eq:ham_mom_d} we can see the long range interactions reduce the coupling between the different states, but do not increase the energy difference between the pure and multiple flow states.

Another difference between this model and that of the SQUID is the setup of the loop. Although our system forms a loop there are only three points at which the atoms can sit. In the SQUID there are sections that are superconducting. In these sections the long range interactions prevent atoms having different flow~\cite{foldy_61}, and make the cat state less phase sensitive. The gap to excited states in the \emph{charged} superconductor is, of course, the plasma frequency.

 It is the junction that acts as a mechanism that creates the superposition of phases of the superfluid wavefunction. This is equivalent to the decaying atom in Schr\"{o}dinger's cat thought experiment. Due to the small numbers of atoms in the junction it is relatively easy to create a superposition of states of uncertain phase, along with the demands of the uncertainty relation $\Delta\phi\Delta N \approx 1$. The phase of the junction is coupled to the rest of the superfluid atoms, because of phase matching around the loop. If the loop contained a macroscopic number of atoms then this would in term form a macroscopic superposition. This is equivalent to the life of the cat being coupled to the atom. 

In a SQUID, if the junction forms a superposition of phases then the rest of the loop will ``follow'', as excitations of the super conductor are limited by the coulomb interaction, which produces a gap equal to the plasma frequency. Future work will examine this issue further to see how a SQUID like systems can be made using neutral atoms. Hopefully this will allow us to understand how to make stable cat states by showing the junctions only require numbers of atoms in the range we have investigated here and that they can be coupled to large numbers of superfluid atoms. Excitations are certainly a problem when the junction is coupled to the neutral superfluid regions. If the superposition is not produced adiabatically then phase and density fluctuations will be produced in the loop. Hopefully atom-atom interactions can be used to limit this effect in the same way that the coulomb force limits excitations in the SQUID~\cite{leggett_87a}.

\section{Conclusion}
We show three reasons, other than decoherence, why macroscopic quantum effects are so hard to see in Bose-Einstein condensates. We illustrate this by finding macroscopic superposition states of flow, firstly in a 1D loop, then in a ring of coupled Bose-Einstein condensates. Firstly, we show that the energy of macroscopic states varies rapidly with the externally applied phase making it hard to keep the degeneracy needed for good coupling between them. Secondly, this coupling decreases as the size of the system increases, due to a larger number of intermediate states being involved. Finally, as more particles are introduced there are more possible flow states for the system to occupy. Even if the amplitudes for being in a multiple flow state are small it still reduces the total amplitude in the two macroscopic states. To avoid this we need a large energy gap between our macroscopic superposition states and the multiple flow states. Our results clearly show why it is hard to make systems with macroscopic quantum properties and will be useful in finding the best setup to produce a macroscopic superposition. These problem can be alleviated by coupling a microscopic mechanism that creates the superposition to a macroscopic system, as in Schr\"{o}dinger's initial thought experiment where the microscopic mechanism is the decaying atom and the macroscopic object is the cat. In the context of the systems studied in this paper, an appropriate solution may be a microscopic Josephson junction coupled to a macroscopic superfluid ring.



\begin{thebibliography}{99}

\bibitem{leggett_02} A.J. Leggett, J. Phys.: Condens. Matter {\bf14} R415 (2002).

\bibitem{schrodinger_35} E. Schr\"{o}dinger, Naturwissenschaften {\bf 23} 807 (1935).

\bibitem{caldeira_81} A.O. Caldeira, A.J. Leggett, Phys. Rev. Lett. {\bf46} 211 (1981).

\bibitem{leggett_87} A.J. Leggett, \emph{et al.}, Rev. Mod. Phys. {\bf59}, 1 (1987).

\bibitem{tonomura_02} A. Tonomura, Physics World (Sept. 2002). 

\bibitem{merli_76} P.G. Merli, G.F. Missiroli, G. Pozzi, Am. J. Phys. {\bf44}, 306 (1976).

\bibitem{tonomura_89} A. Tonomura, J. Endo, T. Matsuda, T. Kawasaki, H. Ezawa, Am. J. Phys. {\bf57} 117 (1989).

\bibitem{carnal_91} O. Carnal, J. Mlynek, Phys. Rev. Lett. {\bf66} 2689 (1991).

\bibitem{Keith_91} D.W. Keith, C.R. Ekstrom, Q.A. Turchette, D.E. Pritchard, Phys. Rev. Lett. {\bf66} 2693 (1991).

\bibitem{Noel_95} M.W. Noel, C.R. Stroud Jr., Phys. Rev. Lett. {\bf75} 1252 (1995).

\bibitem{arndt_99} M. Arndt, O. Nairz, J. Vos-Andreae, C. Keller, G. van der Zouw, A. Zeilinger, Nature {\bf401} 680 (1999).

\bibitem{brezger_02} B. Brezger, L. Hackermüller, S. Uttenthaler, J. Petschinka, M. Arndt, A. Zeilinger, Phys. Rev. Lett. {\bf88} 100404 (2002).

\bibitem{lee_05} J. Lee, A.K. Khitrin, Appl. Phys. Lett. {\bf87} 204109 (2005).

\bibitem{leibfried_05} D. Leibfried \emph{et al.}, Nature {\bf438} 639 (2005).

\bibitem{ourjoumtsev_06} A. Ourjoumtsev, R. Tualle-Brouri, J. Laurat, P. Grangier, Science {\bf312} 83 (2006).

\bibitem{Mitchell2004a} M.W. Mitchell, J.S. Lundeen, and A.M. Steinberg, Nature (London) {\bf 429}, 161 (2004). 

\bibitem{Arndt1999a} M. Arndt {\it et al.} Nature (London) {\bf 401}, 680 (1999).

\bibitem{Sackett2000a} C.A. Sackett {\it et al.} Nature (London) {\bf 404}, 256 (2000).

\bibitem{rouse_95} R.~Rouse, S.~Han, J.E.~Lukens, Phys. Rev. Lett. {\bf 75} (1995) 1614.

\bibitem{friedman_00} J.R. Friedman, V.Patel, W. Chen, S.K. Tolpygo, J.E. Lukens, Nature {\bf406} 43 (2000).

\bibitem{wal_00} C.H. van der Wal, A.C.J. ter Haar, F.K. Wilhelm, R.N. Schouten, C.J.P.M. Harmans, T.P. Orlando, S. Lloyd, and J.E. Mooij, Science {\bf 290} 773 (2000).

\bibitem{jaksch_98} D.~Jaksch, C.~Bruder, J.I.~Cirac, C.W.~Gardiner, P.~Zoller, Phys. Rev. Lett. {\bf 81}, 3108 (1998).

\bibitem{rosenfeld_65} L.~Rosenfeld \emph{Theory of Electrons}, New York: Dover (1965).

\bibitem{saba_05} M.~Saba, T.A.~Pasquini, C.~Sanner, Y.~Shin, W.~Ketterle, D.E.~Pritchard, Science {\bf 307} 1945 (2005).

\bibitem{shin_05} Y.~Shin, G.B.~Jo, M.~Sab, T.A.~Pasquini, W.~Ketterle, D.E.~Pritchard, arXiv:cond-mat/0507154v1 (2005).

\bibitem{jaksch_03} D.~Jaksch, P.~Zoller, New J. Phys. {\bf 5} 56 (2003).

\bibitem{juzeliunas_05} G.~Juzeli\={u}nas, P.~\"{O}hberg, J.~Ruseckas, A.~Klein, Phys. Rev. A {\bf 71} 053614 (2005).

\bibitem{leggett_87a} A.J.~Leggett, \emph{in Chance and Matter}, edited by J. Souletie, J. Vannimenus, and R. Stora (North-Holland, Amsterdam, 1987).

\bibitem{cirac_98} J.I.~Cirac, M.~Lewenstein, K.~M{\o}lmer, P.~Zoller, Phys. Rev. A {\bf 57} 1208 (1998).

\bibitem{gordon_99} D.~Gordon, C.M.~Savage, Phys. Rev. A {\bf 59} 4623 (1999);

\bibitem{ruostekoski_98} J.~Ruostekoski, M.J.~Collett, R.~Graham, D.F.~Walls, Phys. Rev. A {\bf 57} 511 (1998).

\bibitem{dunningham_01} J.A.~Dunningham, K.~Burnett, J. Mod. Opt. {\bf 48} 1837 (2001).

\bibitem{dunningham_06} J.A.~Dunningham, D.W. Hallwood, Phys. Rev. A {\bf 74} 023601 (2006).

\bibitem{hallwood_06} D.W.~Hallwood, K.~Burnett, J.~Dunningham, New J. Phys. {\bf 8} 180 (2006).

\bibitem{rey_03} A.M.~Rey, K.~Burnett, R.~Roth, M.~Edwards, C.J.~Williams, C.W.~Clark, J. Phys. B {\bf 36} 825 (2003).

\bibitem{ashcroft_76} N.W.~Ashcroft, N.D. Mermin, \emph{Solid State Physics}, Saunders College (1976).

\bibitem{bransden_00} B.H. Bransden, C.J. Joachain, \emph{Quantum Mechanics, 2nd edition}, (2000).

\bibitem{goral_02} K.~G\'{o}ral, L.~Santos, Phys. Rev. A {\bf66} 023613 (2002).

\bibitem{marinescu} M.~Marinescu, L.~You, Phys. Rev. Lett. {\bf 81} 4596 (1998).

\bibitem{kim} J.~Kim, B.~Friedrich, D.P.~Katz, D.~Patterson, J.D.~Weinstein, R.~DeCarvalho, J.M.~Doyle, Phys. Rev. Lett. {\bf 78} 3665 (1997).

\bibitem{goral_00} K.~G\/{o}ral, K.~Rz\c{a}\.{z}ewski, T.~Pfau, Phys. Rev. A {\bf 61} 051601 (2000).

\bibitem{santos_00} L.~Santos, G.V.~Shlyapnikov, P.~Zoller, M.~Lewenstein, Phys. Rev. Lett. {\bf 85} 1791 (2000).

\bibitem{goral_02a} K.~G\/{o}ral, L.~Santos, Phys. Rev. A {\bf 66} 023613 (2002).

\bibitem{damski_03} B.~Damski, L.~Santos, E.~Tiemann, M.~Lewenstein, S.~Kotochigova, P.~Julienne, P.~Zoller, Phys. Rev. Lett. {\bf 90} 110401 (2003). 

\bibitem{afrousheh_04} K.~Afrousheh, P.~Bohlouli-Zanjani, D.~Vagale, A.~Mugford, M.~Fedorov, J.D.D.~Martin, Phys. Rev. Lett. {\bf 93} 233001 (2004). 

\bibitem{ronen_06} S.~Ronen, D.C.E.~Bortolotti, D.~Blume, J.L. Bohn, Phys. Rev. A {\bf 74} 033611 (2006).

\bibitem{foldy_61} L.L.~Foldy, Phys. Rev. {\bf 124} 649 (1961).

\end{thebibliography}
\end{document}